\documentclass[rnote]{aa}
\usepackage[varg]{txfonts}

\def\aj{AJ}

\def\apjl{ApJ}
\def\aap{A\&A}
\def\aaps{A\&AS}

\def\mnras{MNRAS}
\def\Msun{M$_\odot$}

\begin{document}
\title{The multiplicity of $\phi$ Phe revisited}
\author{D.~Pourbaix\inst{1}\fnmsep\thanks{Senior Research Associate, F.R.S.-FNRS, Belgium}
\and H.~M.~J.~Boffin\inst{2}
\and R.~Chini\inst{3}\fnmsep\inst{4}
\and T.~Dembsky\inst{3}}
\institute{Institut d'Astronomie et d'Astrophysique, Universit\'e Libre de Bruxelles (ULB), Belgium\\
\email{pourbaix@astro.ulb.ac.be}
\and ESO, Alonso de C\'ordova 3107 Vitacura, Casilla 19001 Santiago, Chile
\and Astronomisches Institut, Ruhr-Universit\"at Bochum, Universit\"atsstrasse 150, 44780 Bochum, Germany
\and Institudo de Astronomia, Universidad Cat\'olica del Norte, Antofagasta, Chile}

\date{Received April, 15, 2013; accepted April, 29, 2013}

\abstract{}
{The chemically peculiar B star $\phi$ Phe was, until very recently, considered a triple system, even though the data were not conclusive and the orbits rather uncertain.  Very recent results by \citet{Korhonen-2013:a} provided a revised orbit, different from the then available astrometric Hipparcos orbit.}
{Additional spectroscopic data, obtained with the BESO spectrograph at Cerro Armazones, confirm the newly found orbit, even though the resulting radial velocities do not allow to improve on the recent orbit. We combine the latter with the Hipparcos measurements to secure the astrometric orbit, and derive the inclination of the system. Using evolutionary tracks, we can finally constrain all the parameters of the two components in this system.}
{We confirm the mass of the primary, 3 M$_\odot$, and find that the companion has a mass of 0.9 M$_\odot$. The inclination of the system is $i=93^{\circ} \pm 4.7^{\circ}$, and is potentially eclipsing; we predict the time of the next conjunction. Given that the eccentricity of the orbit and the exact value of the semi-amplitude of the radial velocity relies on just one set of points, we also urge observers to measure radial velocities at the next periastron passage in April 2015.}
{}

\keywords{astrometry -- (stars)binaries: spectroscopic --techniques:spectroscopy}

\maketitle

\section{Introduction}

\object{$\phi$ Phe} (HIP~8882, HD~11753, HR~558) is a 5th magnitude late-type B star whose chemical peculiarity was first noticed by A.R.~Hyland \citep{Dworetsky-1969:a}.  Its MK classification is B9pHgMn \citep{MKSpCl6}.

It was suggested very early on \citep{Campbell-1928} that the radial velocity of $\phi$ Phe was varying.  However, that claim was only confirmed by \citet{Dworetsky-1982:a} on a dataset too small to yield any orbit, not even the period. \citet{Leone-1999:a} completed the dataset and derived the first spectroscopic orbit ($e=0.32$, $P=41.489$ days).  \citet{Grenier-1999:a} also complemented the dataset with one more radial velocity but did not derive any orbit.  In the original reduction of the Hipparcos data \citep{Hipparcos}, a circular nearly edge-on 878-day orbit was derived from the sole astrometric observations.  $\phi$ Phe thus made it to the triple star class \citep{Tokovinin-2008:a} despite the excellent astrometric fit ($F2=0.10$), leaving no room for any detectable wobble caused by the inner component.  

Even though the spectroscopic orbit has lately been substantially revised \citep{Korhonen-2013:a} with a period now reaching sightly more than three years, the multiplicity of the system was not questioned.  Using the Hipparcos data, we are going to show that $\phi$ Phoenici turns out to be a genuine binary (Sect.~\ref{sect:Astrometry}), thus allowing the system to be fully characterised (Sect.~\ref{sect:Astrophysics}).
 
\begin{table*}[htb]
\caption[]{\label{tab:BESOrv}BESO radial velocities.  Typical uncertainty is 2 km\,s$^{-1}$.}
\centering
\begin{tabular}{llclcc}
\hline\hline
Date & HJD-2\,400\,000 &  Vel (km\,s${^-1}$) & Date & HJD-2\,400\,000 &  Vel (km\,s${^-1}$)\\
\hline
2011/11/09& 55875.52796296 & 6.52 &2012/09/07& 56178.85435184 & 16.13\\
2012/06/22& 56101.90094908 & 10.23&2012/09/16& 56187.67648149 & 13.757\\
2012/06/27& 56106.8688426  & 10.63&2012/09/27& 56198.61953704 & 9.19\\
2012/07/06& 56115.93180557 & 7.36&2012/10/14& 56215.67843749 & 12.93\\
2012/07/15& 56124.86010416 & 10.85&2012/10/16& 56217.65494213 & 13.245\\
2012/08/10& 56150.84261572 & 12.87&2012/10/20& 56221.62182871 & 17.56\\
2012/08/15& 56155.89633103 & 10.15&2012/10/24& 56225.60006944 & 15.30\\
2012/08/29& 56169.83555555 & 13.485&2012/12/04& 56266.68113426 & 15.54\\
2012/09/02& 56173.69356481 & 13.952 \\
\hline
\end{tabular}
\end{table*}

\section{Revised spectroscopic orbit}
Using about 150 Coralie spectra taken in 2000, 2001, 2009, and 2010 together with a few FEROS and HARPS data points on top of the older radial velocities, \citet{Korhonen-2013:a} have recently drastically changed the orbit.  Instead of a month-long period, they obtained a 3-year orbit.  Although these new data strongly constrain the top of the velocity curve, its bottom (and thus partly its semi-amplitude) is still essentially set by the sole six additional data from \cite{Leone-1999:a}.

The Bochum Echelle Spectroscopic Observer (BESO) is a fibre-fed spectrograph based on slightly improved blueprints of the European Southern Observatory FEROS instrument (located at La Silla).  It belongs to the Astronomical Institute of the Ruhr-Universit\"at Bochum (AIRUB) and is located at Cerro Armazones in the Atacama desert in Chile through a partnership between AIRUB and the Universidad Cat\'olica del Norte in Antofagasta.  BESO is attached to the 1.5-m Hexapod Telescope.  It operates in the 370-860 nm wavelength range with an average resolution of 48\,000, varying from order to order between 43\,000 and 60\,000 \citep[see][for more details about BESO]{Fuhrmann-2011:a}.

For purely astrometric reasons, $\phi$ Phe has been monitored by BESO since late 2011.  The initial goal of that monitoring was to secure the spectroscopic orbit of the inner components (even the period was still uncertain after \citet{Leone-1999:a}) which could then be used to improve the Hipparcos solution.  Seventeen spectra were obtained between November 2011 and October 2012 yielding the radial velocities listed in Tab~\ref{tab:BESOrv}. The radial velocities were determined both by fitting some of the strongest lines and by cross-correlation using the xcsao task in IRAF. Depending on the quality of the spectrum - the S/N varies between ~30 to ~70 - the formal errors of the cross-correlation is between 0.3 and 1 km\,s$^{-1}$. The real error is higher as it combines systematics from the wavelength calibration and intrinsic variability of the object as reported by \citet{Korhonen-2013:a}.  The resulting rms for these sole data is 1.92 km\,s$^{-1}$, likely caused by the BESO calibration itself which was reported to have had some problems during our run.  At the time our investigation on $\phi$ Phe was initiated, the results from \citeauthor{Korhonen-2013:a} had not been released and the adopted S/N was thought to be enough to assess the 41-day orbit.  However, unlike all the other existing datasets, this one covers 4 consecutive months (as only one measurement was taken in November '11 before $\phi$ Phe disappeared for 6 months).

\begin{figure*}[ht]
\begin{center}
\resizebox{0.49\hsize}{!}{\includegraphics{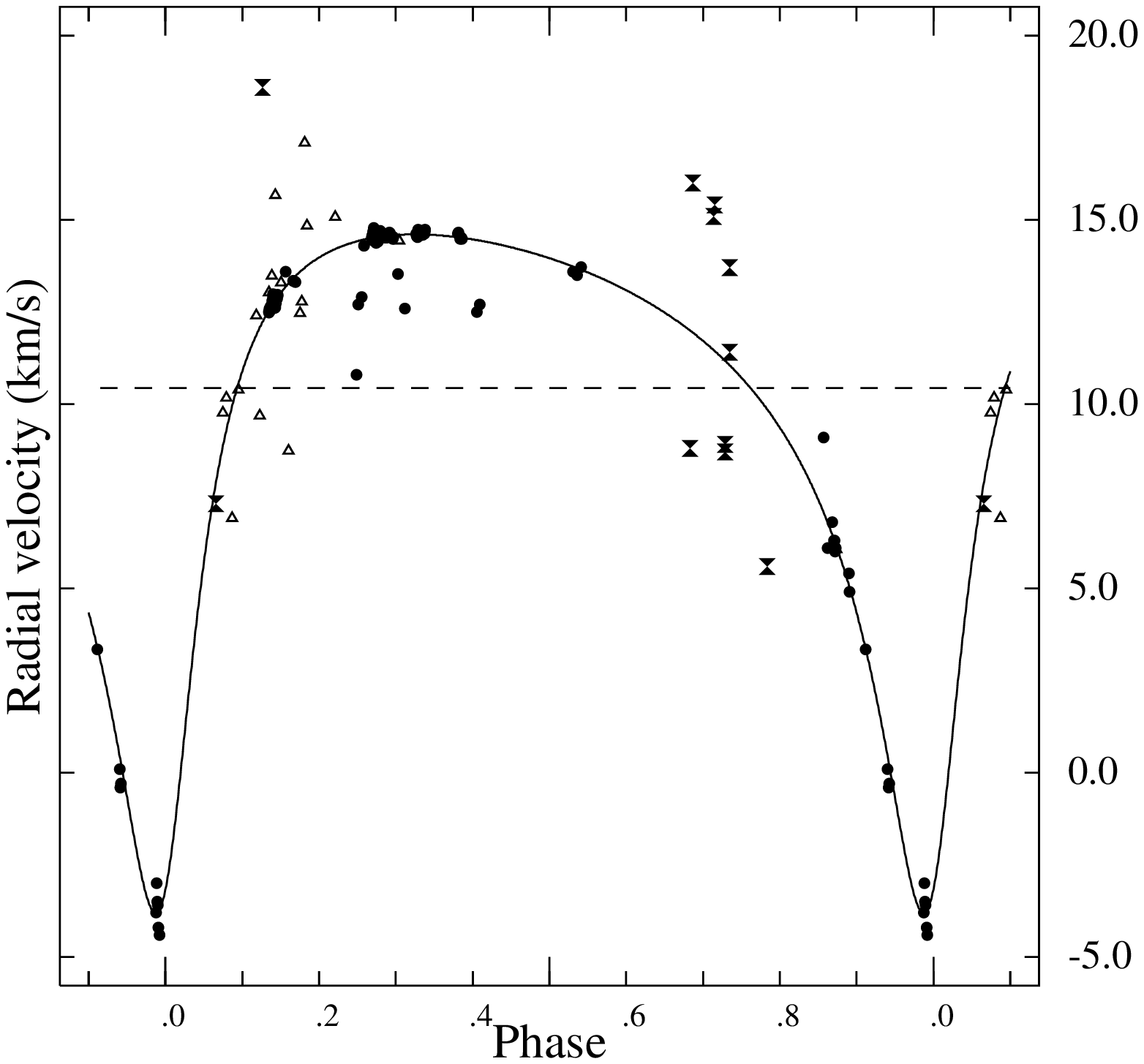}}
\resizebox{0.49\hsize}{!}{\includegraphics{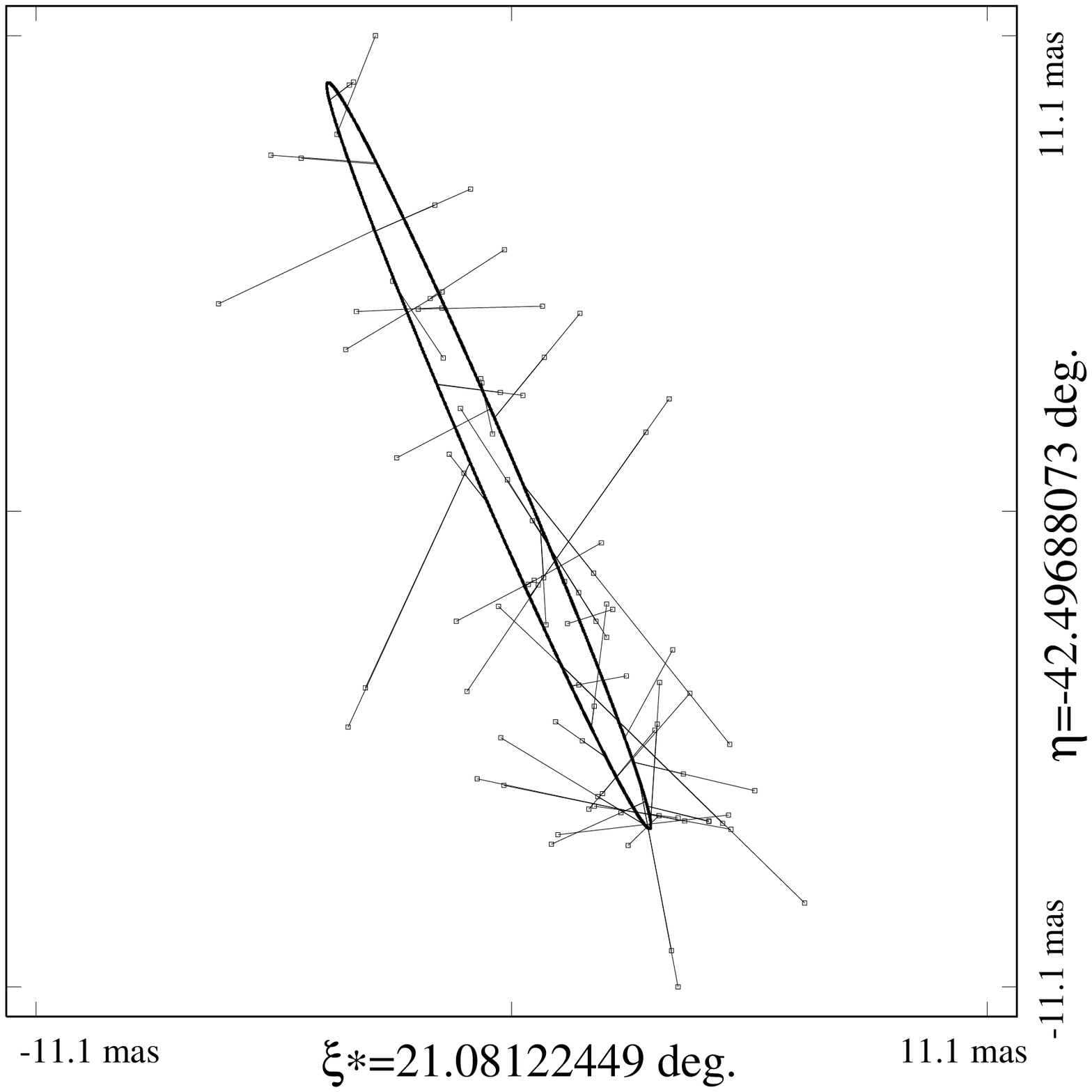}}
\end{center}
\caption[]{\label{fig:RevisedOrbits}{\bf Left panel:} Spectroscopic orbit with data from \citet{Leone-1999:a}, \citet{Grenier-1999:a}, \citet{Korhonen-2013:a} (all as filled disks), BESO (open triangles) and \citet{Campbell-1928} as filled diabolos.  {\bf Right panel:} New astrometric orbit based on the original Hipparcos observations.}
\end{figure*}

Despite the extended phase coverage, the precision of these BESO radial velocities is not good enough to substantially refine the orbit from \citet{Korhonen-2013:a}.  However, given their continuity, our data points definitively exclude any 40-day period orbit as proposed by \citet{Leone-1999:a}.  The orbit from \citeauthor{Korhonen-2013:a} is plotted in the left panel of Fig.~\ref{fig:RevisedOrbits}.  The scatter of both the BESO data and the old velocities from \cite{Campbell-1928} make the usefulness of these sets rather limited, especially the latter as far as constraining the period is concerned.

\section{Astrometric counterpart}\label{sect:Astrometry}

The original choice of adopting a circular orbit to fit the Hipparcos data of this object was for the sake of simplification.  Indeed, it means that not only the eccentricity but also the argument of the periastron ($\omega$) can be set to 0.  That orbit is listed in the first column of Tab.~\ref{tab:AstrometricOrbits}.  Strangely enough, the very same orbit was also adopted in the new reduction of the Hipparcos observations by \citet{Hip2} despite the seemingly poor fit resulting from that choice: 0.1 versus 12.34 for the goodness of fit (Wilson \& Hilferty's cube root transformation which follows a $N(0,1)$ distribution \citep{KeAdThSt1}).  However, in this second reduction of HIP~8882, there was no new orbit fitting per se, the original orbit was simply adopted.

In their general attempt to reprocess the Hipparcos observations of the spectroscopic systems listed in $S_B^9$ \citep{Pourbaix-2004:b}, \citet{Jancart-2005:a} concluded that the orbit by \citet{Leone-1999:a} was not present in the astrometric data. Neither the spectroscopic orbit nor the astrometric one were ever questioned despite the absence of hint of the latter in the radial velocities.

The astrometric orbits resulting from both the Campbell and the Thiele-Innes' approach are listed in Tab.~\ref{tab:AstrometricOrbits}.  The assessment of such spectro-astrometric combinations was extensively described and improved in several papers \citep{Pourbaix-2001:b,Pourbaix-2003:a,Jancart-2005:a}.  It essentially consists in fitting the astrometric data with two different approaches (freezing some parameters in one while leaving them free in the other) and in evaluating the consistency between these two solutions through several statistical tests.  In both cases, the eccentricity, period and periastron time are adopted from the spectroscopic orbit.  In Campbell's approach, the amplitude of the radial velocity curve and the argument of the periastron are also adopted from the spectroscopic side.  In the alternative, the four Thiele-Innes constants are left free.  In no case, the radial velocities are fitted so the $K_1$ listed in Tab.~\ref{tab:AstrometricOrbits} is derived from the astrometric solution alone.  If both solutions are consistent (which is the case here), Campbell's one should be favoured as it results from the larger number of degrees of freedom.

\begin{table*}[htb]
\caption[]{\label{tab:AstrometricOrbits}Astrometric orbits based on the original Hipparcos data with and without spectroscopic help.  The second column lists the published solution \citep{Hipparcos}.  As in the Hipparcos Catalogue, $F2$ denotes the Goodness of Fit and $\varpi$ the parallax.  The third column lists the proper motion after Tycho-2 \citep{Hog-2000:a}.}
\centering
\begin{tabular}{lcccc}
\hline\hline
        &          & & \multicolumn{2}{c}{Spectro-based}\\
Element & Original & Tycho-2 & Campbell & Thiele-Innes \\
\hline
$e$     & 0 (fixed) & & \multicolumn{2}{c}{0.589 (fixed)} \\
$P$ (d) & $877.6541\pm117.3686$ & & \multicolumn{2}{c}{1126.11 (fixed)} \\
$T_0$ (JD-2\,400\,000)  & $48\,589.2621\pm38.5598$ && \multicolumn{2}{c}{53\,766.2 (fixed)} \\ 
$a_0$ (mas) & $5.91\pm0.93$ & & $8.4\pm0.52$& 9.8 \\
$i$ (deg.) & $93.51\pm5.34$ & & $93\pm4.7$& 93\\
$\omega$ (deg.) & 0 (fixed) & & 201.6 (fixed)& 196\\
$\Omega$ (deg.) & $161.88\pm6.94$ & & $159\pm6.4$& 156\\
$A$ (mas) & & & & $8\pm1.0$ \\
$B$ (mas) & & & & $-4\pm1.0$ \\
$F$ (mas) & & & & $-2.7\pm0.80$ \\
$G$ (mas) & & & & $-0.6\pm0.67$ \\
$F2$      & 0.10 & & -0.83 & -0.87\\
$K_1$ (km\,s$^{-1}$) & 6.93 & & 9.21 (fixed)& 11\\
\vspace{0.5em}\\
$\varpi$ (mas) & $10.55\pm0.69$ & & $10.87\pm0.67$ & $10.6\pm0.69$\\
$\mu_{\alpha*}$ (mas\,yr$^{-1}$)  &$-34.32\pm0.68$ & $-33.7\pm0.7$ & $-35.3\pm0.74$ & $-35.7\pm0.78$\\
$\mu_{\delta}$ (mas\,yr$^{-1}$)  &$-28.34\pm1.39$ & $-26.0\pm0.6$& $-25.8\pm0.52$& $-25.4\pm0.61$\\
\hline
\end{tabular}
\end{table*}

All the assessment indicators are listed in Tab.~\ref{tab:hipAssessment} and confirm the excellent agreement between the spectroscopic orbit and its astrometric counterpart.  The goodness of fit is significantly reduced with the eccentric orbit with respect to the original circular one.  Despite the change of the eccentricity and period, the parallax and the proper motion, last three lines of Tab.~\ref{tab:AstrometricOrbits}, remain unchanged (within the error bars).

\begin{table}[ht]
\caption[]{\label{tab:hipAssessment}Assessment of the presence of the spectroscopic orbit in the Hipparcos observations.  The notations are those from \citet{Pourbaix-2003:a}}
\centering
\begin{tabular}{lclc}
\hline\hline
Indicator & Value & Indicator & Value\\
\hline
$Pr_1$ & $<10^{-6}$ & $\epsilon$ & 0.589825\\
$Pr_2$ & $<10^{-6}$ & $D$ & 1.922401\\
$Pr_3$ & $<10^{-6}$ & $Pr_4$ & 0.645796\\
$F2_{\rm TI}$ & -0.866067 & $Pr_5$ & 0.408080\\
\hline
\end{tabular}
\end{table}

Although the Thiele-Innes solution fits the Hipparcos data very well, the resulting $K_1$ is slightly larger than the spectroscopic value of \citet{Korhonen-2013:a}, i.e. 9.21 km\,s$^{-1}$.  An astrometric value smaller than its spectroscopic counterpart is often the sign that the assumption about the absence of light from the secondary is questionable (very unlikely, as we will see at the end of Sect.~\ref{sect:Astrophysics}).  In the case of HIP~8882, could the spectroscopic value be underestimated?  This cannot be ruled out as the bottom of the velocity curve is constrained by the six points over five consecutive days from \citet{Leone-1999:a}, with the observed velocity ranging from -4.4 to -3.0 km\,s$^{-1}$.  The next minimum of the radial velocity will take place on JD 2\,457\,130 (2015.290).

For the sake of completeness, we did fit the data coming from the second reduction of the Hipparcos observations \citep{Hip2}.  The resulting $a_0$ is way too small ($4.20\pm0.33$ mas) and, even though the goodness of fit goes from 12.34 to 5.39, all the statistical indicators are made useless by the poor weighting scheme adopted in that reduction.  It is therefore safer not to use it, at least for binaries.

Although a perfectly edge-on orbit would be consistent with the astrometric solution, there is no sign of any eclipse ever reported.  The Hipparcos photometry \citep{Hipparcos} shows a constant brightness during the whole mission.  The difference between the 5th and 95th percentiles is 0.02 Hp mag and there is no sign of brightness decrease at the epoch of conjunction.

Finally, the revision of the orbit leads to a significant change in the proper motion with respect to the Hipparcos one, especially along the declination axis, getting much closer to the Tycho-2 value \citep{Hog-2000:a}.  Unfortunately, the right ascension component moves in the opposite direction.  However, the UCAC4 \citep{Zacharias-2013:a} proper motion $(-33.9\pm1.0;-28.2\pm1.0)$mas\,yr$^{-1}$ confirms neither the Tycho-2 one nor ours, illustrating once more our poorly known this bright system is.

\section{Astrophysical outcome}\label{sect:Astrophysics}
Besides the genuine binary nature of $\phi$ Phe, our new combined astrometric-spectroscopic solution confirms the parallax of the system and the orbital inclination, which is now safe to combine with the mass function to derive the mass of the secondary.

\begin{figure*}[ht]
\begin{center}
\resizebox{0.49\hsize}{!}{\includegraphics{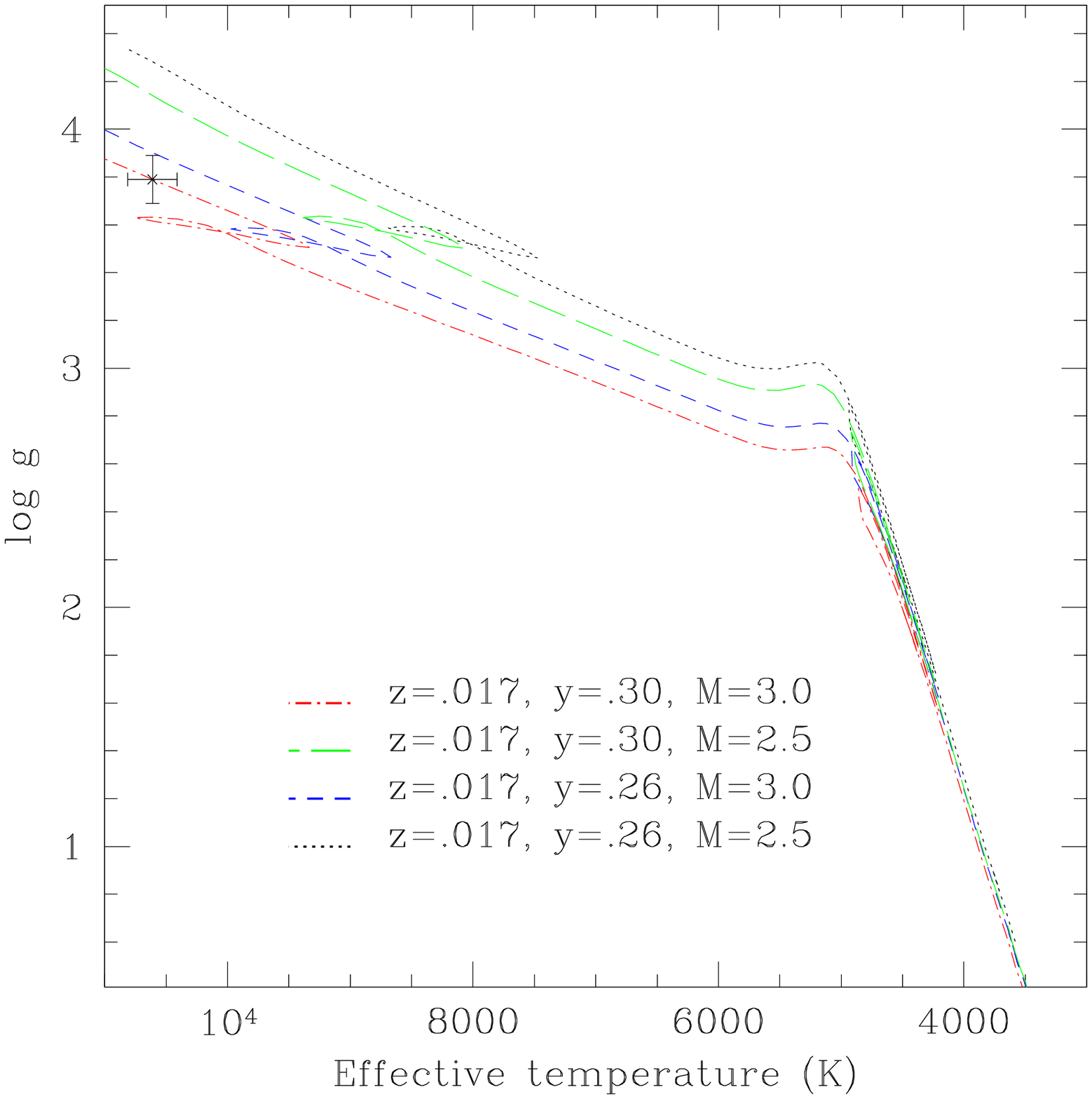}}
\resizebox{0.49\hsize}{!}{\includegraphics{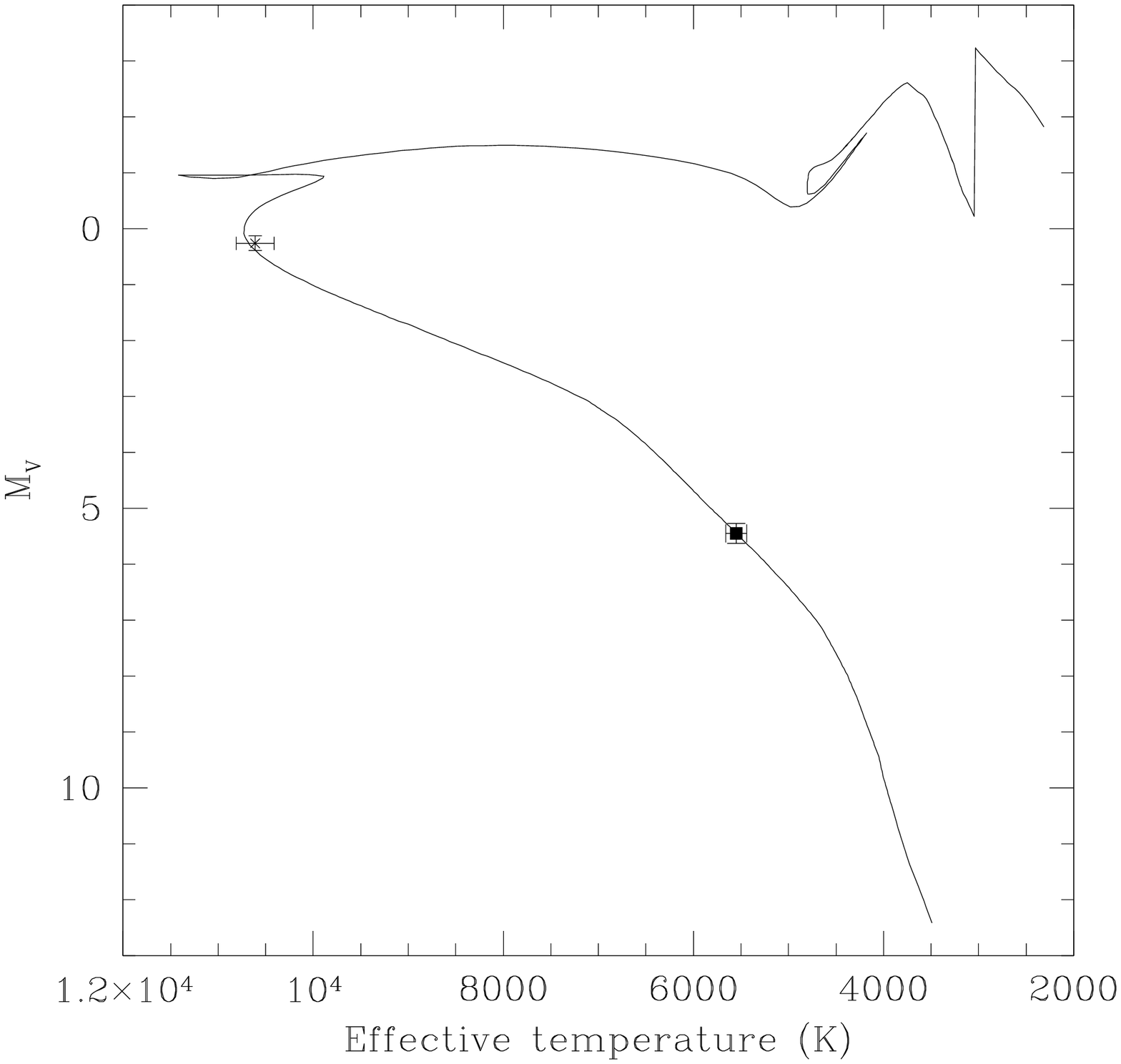}}
\end{center}
\caption[]{\label{fig:iso}{\bf Left panel:} Four evolutionary tracks from \citep{Bertelli-2009:a}.  {\bf Right panel:} $\log {\rm age(yr)}=8.41$-isochrone with same $Y$ and $Z$ as in the best evolutionary track.}
\end{figure*}

\citet{Dolk-2003:a} reported $T_\mathrm{eff}=10,612\pm200$ K and $\log g=3.79\pm0.10$ together with ${\rm [Fe/H]}=0.15$ from \citet{Smith-1993:a}.  On the grid of stellar evolutionary tracks from \citet{Bertelli-2009:a}, the 3-solar mass track with $Y=0.3$ and $Z=0.017$ gives the best match as far as $\log g$ and $T_\mathrm{eff}$ are concerned (left panel of Fig.~\ref{fig:iso}).  According to \citet{Hakkila-1997:a}, the interstellar extinction $A_V=0.03\pm0.17$ which coupled to $V=5.112$ yield the absolute magnitude $M_V=0.26\pm0.13$.  The point $(T_\mathrm{eff},M_V)$ is in excellent agreement with the $\log {\rm age(yr)}=8.41$-isochrone from \citep{Bertelli-2009:a} as illustrated in the right panel of Fig.~\ref{fig:iso}.

From the spectroscopy, $f(m)=0.048\pm0.0015$\Msun, which, combined with $i=93^{\circ}$ (astrometry) and $M_1=3.0\pm0.12$\Msun\ (evolutionary track), lead to $M_2=0.91\pm0.025$\Msun.  Using one solar mass as initial guess for $M_2$, the dynamical parallax method \citep{PrDoSt} yields $M_2=0.9045$\Msun\ after a few iterations, very consistent with our track estimate.  On the $\log {\rm age(yr)}=8.41$-isochrone (Fig.~\ref{fig:iso}), such a mass corresponds to the big point on the right panel.  The magnitudes of the two components differ by 5.7 in $V$ and 3.9 in $K$.  Such a large difference in $V$ explains the absence of signal in the Hipparcos data and in the spectra.  The semi-major axis of the relative visual orbit is 36.3 mas only, making this system essentially inaccessible to VLT/NACO \citep{Scholler-2010:a}.  The two components are nevertheless bright enough to be resolved with the VLTI.  

Even though, with $\Delta m=5.7$ in $V$, some hints of the secondary peak might be detectable in FEROS-based cross correlation function, a thorough inspection of the publicly available spectra has not revealed anything.  Should eclipses occur (JD 2\,457\,078, i.e. 2015.148), they would yield a change in magnitude of 0.010 and 0.057 in $V$ and $K$ respectively, so the constant brightness reported by Hipparcos cannot be used to rule out the possibility of eclipses.  Such eclipses would also constrain the orbit to be edge-on within $0.0001^{\circ}$.

\section{Conclusion}
It is rather disappointing to note that despite the fact that $\phi$ Phe belongs to the Bright Star Catalogue, in which it was noted for its chemical peculiarities, one has to wait till 2013 to have the first plausible characterisation of the components of this binary system.

Out of the 235 orbital solutions published in Hipparcos \citep{Hipparcos}, 115 were derived without using any ground-based solution, 70 of which had their eccentricity and periastron angle arbitrarily set to 0 as for $\phi$ Phe.  How many other systems deserve some revision?

\begin{acknowledgements}
We thank the referee, A.~Tokovinin, for his useful comments and J.F.~Gonz\'alez and M.~Briquet for sending the radial velocities from \citet{Korhonen-2013:a} prior to their upload to the CDS.  This publication is supported as a project of the Nordrhein-Westf\"alische Akademie der Wissenschaften und der K\"unste in the framework of the academy program by the Federal Republic of Germany and the state Nordrhein-Westfalen.  This research has made use of the Simbad data base, operating at CDS, Strasbourg, France.
\end{acknowledgements}

\end{document}